%% file: main.tex
\documentclass[sigconf,nonacm]{acmart}

\usepackage{subcaption}
\usepackage{rotating, graphicx}

\usepackage{booktabs}
\usepackage{tabularx}
\usepackage[normalem]{ulem}
\useunder{\uline}{\ul}{}
\usepackage[export]{adjustbox}

\usepackage{enumitem}
\usepackage{color}

\definecolor{pblue}{rgb}{0.13,0.13,1}
\definecolor{pgreen}{rgb}{0,0.5,0}
\definecolor{pred}{rgb}{0.9,0,0}
\definecolor{pgrey}{rgb}{0.46,0.45,0.48}

\usepackage{listings}
\usepackage[ruled,vlined]{algorithm2e}

\usepackage{tcolorbox}
\usepackage{mdframed}
\newmdenv[innerlinewidth=0.5pt, roundcorner=4pt,linecolor=gray,innerleftmargin=4pt,
innerrightmargin=4pt,innertopmargin=4pt,innerbottommargin=4pt]{note}

{\medskip\begin{note}\em}%
{\end{note}\medskip}

\author{Milos Ojdanic}
\email{milos.ojdanic@uni.lu}
\affiliation{%
  \institution{University of Luxembourg}
}
\author{Aayush Garg}
\email{aayush.garg@uni.lu}
\affiliation{%
  \institution{University of Luxembourg}
}
\author{Ahmed Khanfir}
\email{ahmed.khanfir@uni.lu}
\affiliation{%
  \institution{University of Luxembourg}
}
\author{Renzo Degiovanni}
\email{renzo.degiovanni@uni.lu}
\affiliation{%
  \institution{University of Luxembourg}
}
\author{Mike Papadakis}
\email{michail.papadakis@uni.lu}
\affiliation{%
  \institution{University of Luxembourg}
}
\author{Yves Le Traon}
\email{yves.letraon@uni.lu}
\affiliation{%
  \institution{University of Luxembourg}
}

\AtBeginDocument{%
  \providecommand\BibTeX{{%
    \normalfont B\kern-0.5em{\scshape i\kern-0.25em b}\kern-0.8em\TeX}}}

\begin{document}

\title{Syntactic Vs. Semantic similarity of Artificial and Real Faults in Mutation Testing Studies}

\begin{abstract}
Fault seeding is typically used in controlled studies to evaluate and compare test techniques. Central to these techniques lies the hypothesis that artificially seeded faults involve some form of realistic properties and thus provide realistic experimental results. In an attempt to strengthen realism, a recent line of research uses advanced machine learning techniques, such as deep learning and Natural Language Processing (NLP), to seed faults that look like (syntactically) real ones,  implying that fault realism is related to syntactic similarity. This raises the question of whether seeding syntactically similar faults indeed results in semantically similar faults and more generally whether syntactically dissimilar faults are far away (semantically) from the real ones. We answer this question by employing 4 fault-seeding techniques (PiTest - a popular mutation testing tool, IBIR - a tool with manually crafted fault patterns, DeepMutation - a learning-based fault seeded framework and CodeBERT - a novel mutation testing tool that use code embeddings) and demonstrate that syntactic similarity does not reflect semantic similarity. We also show that 60\%, 47\%, 43\% and 7\% of the real faults of Defects4J V2 are semantically resembled by CodeBERT, PiTest, IBIR and DeepMutation faults. We then perform an objective comparison between the techniques and find that CodeBERT and PiTest have similar fault detection capabilities that subsume IBIR and DeepMutation, and that IBIR is the most cost-effective technique. Moreover, the overall fault detection of PiTest, CodeBERT, IBIR and DeepMutation was, on average, 54\%, 53\%, 37\% and 7\%.

\end{abstract}

\maketitle

\section{Introduction}
Fault seeding techniques, such as mutation testing, are extensively used in controlled studies to evaluate and compare testing techniques \cite{PapadakisK00TH19, PapadakisHHJT16}. These techniques allow researchers to seed faults under experimentally controlled conditions and thus perform reproducible test assessments. In a sense by comparing the number of seeded faults revealed by test methods, researchers can form a proxy metric that approximates the fault revealing potential of the performed testing \cite{PapadakisSYB18, JustJIEHF14, AndrewsBL05}. 

Although popular, such techniques have been criticised for producing unrealistic faults \cite{SemSeed, tufano2019learning, Wild_Caught, mutations_how_close_are_they_to_bugs}, i.e., faults that are significantly different from real ones in terms of syntax \cite{mutations_how_close_are_they_to_bugs}, and as a result numerous propositions have been made claiming to produce seeded faults that are syntactically similar to real ones. The most recent research in particular, motivated by the code naturalness hypothesis \cite{HindleBSGD12}\footnote{Naturalness hypothesis states that programs exhibit properties similar to text and thus, natural language process techniques can be used to support code analysis techniques.}, aims at forming realistic faults that are, in fact, artificial faults that have some form of syntactic similarity to real ones, i.e., usually following particular syntactic fault patterns. We call this line of work as \textit{fault mimicking} approaches.

Table \ref{tbl:Related} lists a set of recent fault mimicking techniques that aim at generating (syntactically) realistic faults. By inspecting the table it becomes evident that the key objective of these techniques is to improve the realism of fault-seeding, which is defined and evaluated by some form of syntactic distance (number of tokens changed, Bleu score, etc.) from real faults.  While such approaches may indeed succeed to generate some exact matches of real faults, they have never been evaluated with respect  to their semantics and particularly their utility as fault-based testing techniques \cite{PapadakisK00TH19}, in a sense the typical use case of mutation testing \cite{demillo1978, PapadakisK00TH19}.

Mutation testing has long  been based on the basis that fault seeding should be based on untargeted program syntactic changes \cite{offut1992, Offutt:Mutation:2001}. These changes are defined using the programming language grammar and are completely unaware of any fault semantics. The key assumption here is that simple syntactic changes result in semantic deviations that subsume the faulty ones \cite{offut1992}, meaning that test cases revealing these faults are also capable of revealing real faults \cite{ChekamPTH17}. Therefore, the aim of mutation testing is different from the one followed by fault mimicking techniques. 

The key strategy followed by fault mimicking is to identify program locations where fault opportunities emerge and perform relevant changes, following a pattern observed in some fault instances, that alter the program behaviors similarly to real faults. This implies an underlying assumption that \textit{seeding faults with frequent syntactic fault patterns that have similarities with a real fault will result in faults that are subtle or semantically similar to real ones}. Similarly, another assumption is that \textit{seeding faults that are syntactically dissimilar to real ones results in unrealistic faulty semantics}, i.e., the seeded fault semantics are quite different from those of real faults.

These assumptions may appear intuitive but have absence of evidence, except of course from the case where seeded faults match exactly real ones. Early research on the coupling effect \cite{offut1992} stated that ``simple faults can cascade or couple to form other emergent faults'', implying that fault instances couple independent of their pattern. Additionally, recent studies report large semantic overlaps between simple and complex faults \cite{PapadakisHHJT16, KurtzAODKG16, jia2009}, questioning on the role of the syntactic-based metrics. 

This raises the question of whether syntactically similar, or dissimilar, faults are also semantically similar, or dissimilar. More generally, a question of whether the use of such techniques results in faults that: \textit{a) are semantically similar to real faults, b) resemble (semantically) more faults than the dissimilar ones, and c) are subsumed by simple untargeted syntactic deviations as done by mutation testing, i.e., bring any useful addition to mutation testing}.  

 We answer the above questions by employing four fundamentally different fault-seeding techniques. These include PiTest \cite{ColesLHPV16}, a popular mutation testing tool \cite{KintisPPVMT18}, that uses simple syntactic patterns, IBIR \cite{khanfir2020ibir}, a mutation testing tool with manually crafted fault patterns, DeepMutation \cite{DeepMutation}, a deep learning-based tool that derives patterns from real bug-fixes \cite{tufano2019learning}, and CodeBERT \cite{DBLP:conf/emnlp/FengGTDFGS0LJZ20}, a newly designed mutation testing tool that uses a pre-trained language model capable of making token replacement based on big code learning, i.e., code embeddings. Hence, we investigate the ability of all faults produced by these techniques to form similar semantic deviations as the Defects4J V2 faults \cite{defects4j} and check their potential utility within mutation testing. 
 
Perhaps surprising, our results show that syntactic similarity does not reflect semantic similarity indicating that syntactic distance should not be used as an evaluation metric in the context of mutation testing. Additionally, our results show that 60\%, 47\%, 43\% and 7\% of the real faults of Defects4J V2 can be semantically resembled by CodeBERT, PiTest, IBIR and DeepMutation faults, respectively, indicating that fault mimicking, i.e., IBIR and DeepMutation, approaches are significantly outperformed by syntactic-based fault seeding ones, i.e., CodeBERT and PiTest. To further demonstrate this difference, we also perform an objective comparison between the techniques and show that CodeBERT and PiTest have similar fault detection capabilities that subsume IBIR and DeepMutation. We also compare the techniques in an cost-effective way (by controlling test size) and show that when selecting few test cases, IBIR is the best cost-effective solution, followed by CodeBERT and PiTest. DeepMutation is outperformed by the other techniques. In contrast, when selecting sufficient tests to kill more mutants, the fault detection of PiTest and CodeBERT is increasing and reaches  54\% and 53\% (while IBIR and DeepMutaiton reaches 37\% and 7\%).

Overall, our work aims at raising the awareness over the use of non-semantic metrics in fault seeding studies and in providing evidence related to the utility of recent fault seeding advances in software testing. We also 
provide evidence that fault  coupling makes syntactically dissimilar faults to be useful in software testing. 

In summary, our key contributions are the following:

\begin{itemize}
    \item We perform the first study on comparing syntactic and semantic similarity of artificial and real faults and found no link between these two metrics.  
    \item We propose a new fault seeding technique based on a state-of-the-art pre-trained language model, CodeBERT, and show its ability to semantically resemble real faults.  
    \item We show that IBIR is the most cost-effective technique but subsumed by PiTest in almost all cases when not considering cost. We also find that CodeBERT faults can complement the fault set of PiTest, motivating research that exploits the link between mutation testing and code embeddings, and that DeepMutation is almost subsumed by the other techniques.  
\end{itemize}

\begin{table}[]
\vspace{-0.8em}
\centering
\caption{Fault Mimicking Techniques}
\vspace{-1.1em}
\begin{tabular}{p{.107\textwidth}p{.2\textwidth}p{.11\textwidth}}
\toprule
\textbf{\scriptsize{Approach}} & \textbf{\scriptsize{Aim}} & \textbf{\scriptsize{Evaluation metric}}\\ 
\midrule
\scriptsize{SemSeed: Token Embeddings\cite{SemSeed}} & \scriptsize{Derive syntactic patterns that are syntactically similar to real faults} & \scriptsize{Exact syntactic match}\\
\scriptsize{DeepMutation: Learning-based Mutations\cite{DeepMutation}} & \scriptsize{Produce mutants syntactically similar to real faults} & \scriptsize{Syntactic distance from real faults} \\
\scriptsize{Learning-based Mutations\cite{tufano2019learning}} & \scriptsize{Derive syntactic patterns from bug-fixes} & \scriptsize{Syntactic distance from real faults} \\
\scriptsize{Wild-Caught Mutations\cite{Wild_Caught}} & \scriptsize{Deriving simple syntactic patterns from bug-fixes}  & \scriptsize{Token similarity, Compilability} \\ 
\scriptsize{Mutation Monkey\cite{Mutation_Monkey}} & \scriptsize{Deriving common fault syntactic patterns} &  \scriptsize{Detection Ratio} \\ 
\scriptsize{Analysis of real faults and mutants \cite{mutations_how_close_are_they_to_bugs}} &  \scriptsize{Syntactic similarities of bug-fixes and mutants} & \scriptsize{Number of tokens changed} \\
\bottomrule
\end{tabular}
\vspace{-1.1em}
\label{tbl:Related}
\end{table}

\input{sections/background.tex}
\input{sections/motivating_example.tex}

\input{sections/research_questions.tex}
\input{sections/tools.tex}

\input{sections/experimental_setup.tex}
\input{sections/empirical_evaluation.tex}
\input{sections/discussion.tex}

\section{Related Work}

Fault seeding and particularly mutation testing, is widely used in experimental studies as a way to compare and assess testing techniques \cite{PapadakisK00TH19}. Assuming that seeded faults include properties that are in some sense similar to real ones \cite{AndrewsBL05}. Interestingly, mutation testing, one of the most widely used techniques \cite{PapadakisK00TH19}, introduces faults that are syntactically simple and are quite different from real faults that are in their majority more complex \cite{mutations_how_close_are_they_to_bugs}. In particular, the study of Gopinath et al. \cite{mutations_how_close_are_they_to_bugs} provided empirical evidence showing the misalignment between seeded and real faults that are produced by traditional mutation operators and concluded that real faults are rarely equivalent to mutant-faults.

To deal with this issue, Brown et al. \cite{Wild_Caught} proposed inferring fault seeding patterns, mutation operators, by using historical fault-fixing commits. The idea was to form (syntactic) fault patterns that resemble (in terms of syntax) real historical faults. Their results show that syntactic fault patterns can be mined from code versioning systems, and these differ (syntactically) from those used by modern mutation testing tools.  

DeepMutation \cite{DeepMutation}, a neural machine translation technique \cite{tufano2019learning} that automatically infers fault patterns from historical fault-fixing commits, was proposed. It was shown that DeepMutation resembles exact matches of 45\% of real faulty cases while achieving relatively good syntactic similarity scores in most of the cases. SemSeed \cite{SemSeed} aims at inferring faulty patterns from bug-fixes and attempts to generalize them by appropriately adapting them to the particular local code, i.e., context. Although powerful, SemSeed operators on JavaScript programs making its application in our experiment hard. 

More recently, mutation monkey \cite{Mutation_Monkey} was built by mining frequently occurring faults from complex changes that caused operational issues at Facebook \cite{Mutation_Monkey}. The analysis of these faults indicated that they were good at finding holes and missing tests in the systems under test. 
Interestingly, the above studies aim at mimicking (syntactically) real faults, and as a result, they have been evaluated with ``static'' syntactic-nature metrics such as the syntactic similarity. Hence, raising the question of whether they are suitable for dynamic analysis, such as mutation testing, i.e., incorporating realistic semantic fault properties, and how they compare with traditional mutation testing, which we investigate here. 

Traditionally mutation testing aims at seeding faults using simple syntactic changes. Showing empirical evidence of the coupling effect, \cite{demillo1978} states that simple faults subsume almost all the complex ones \cite{offut1992}. This implies a more general assumption about the ''size''  of faults \cite{OffuttH96}, suggesting that seeded faults with small syntactic distance from the original program introduce small semantic deviations (subtle faults), which form valuable test requirements \cite{ChekamPCT21} and lead to high fault reveling potential \cite{PapadakisCT18}.

The coupling between seeded faults has also been considered as a source of bias in mutation testing studies as it introduces large overlaps between the seeded fault instances \cite{jia2009, KurtzAODKG16, PapadakisHHJT16}. Nevertheless, the question of how to select optimal mutant-fault sets falls outside the scope of this work. 


\section{Conclusion}
We investigated the link between syntactic and semantic similarity of seed and real faults in the context of mutation testing. Our results showed that many seeded faults behave similarly to real ones (they have high semantic similarity), while at the same time having low syntactic similarity (to real faults). This means that we found no evidence suggesting any link between syntactic and semantic similarity, except of course from the case of exact matches. 
We also found that CodeBERT can complement PiTest, i.e., the joint use of PiTest and CodeBERT resulted in 25\% more subsuming mutants than PiTest alone, and that IBIR forms the most cost-effective technique. Interestingly, the ability of CodeBERT indicates a potentially new direction for mutation testing research, one that exploits the link between mutation testing and code embedding.

\bibliographystyle{ACM-Reference-Format}
\bibliography{bibfile}

\end{document}

%% file: sections/background.tex
\section{Syntactic and Semantic similarity of Artificial and Real Faults}
\label{mutation_testing}

Mutation is a test criterion in which test requirements are characterized by mean of seeded faults that are called \emph{mutants}. Thus, mutants are artificial faults generated by performing simple syntactic modifications to the program under analysis \cite{PapadakisK00TH19}. For instance, in expressions like  $\texttt{a < b}$ faults  are  seeded by changing (mutating) the expression to  the following one $\texttt{a > b}$. 
Mutant faults are then used to assess the effectiveness and thoroughness of a test suite in detecting these artificial faults. A test case that detects a mutant fault, i.e., it is capable of producing distinguishable observable outputs between the mutant and the original program is said to be able to \emph{kill} the mutant. A mutant is said \emph{killed} if it is detected by a test case or a test suite, otherwise it is called \emph{live} or \emph{survived}. Test adequacy is called \emph{mutation score},  and is computed as the ratio of killed mutants over the total number of generated mutants. 

To evaluate fault seeding two types of metrics are used, the syntactic and semantic similarity. Table \ref{tbl:Related} lists studies using syntactic similarity, while some form of semantic similarity has been used by several studies \cite{jia2009, JustJIEHF14, PapadakisSYB18, khanfir2020ibir}. Intuitively, \emph{syntactic similarity} refers to the distance between the text representations of the mutant and the real faulty code, while \emph{semantic similarity} to the program \emph{behavior} similarities, between the seeded and the real fault.

To compute syntactic similarity between two sequence of tokens we consider 
Bilingual Evaluation Understudy (BLEU) score~\cite{metrics_bleu} which is widely used for quantifying machine translated text in NLP~\cite{bleu_word_string_similarity, bleu_seq2seq, bleu_phrase_translation, bleu_exploring_tl}. 
BLEU score takes a reference and a candidate text, brakes it into n-grams and computes how many n-grams of the candidate text appear in the reference text. We report the geometric mean of all n-grams up to 4, similarly to previous work~\cite{tufano2019learning}.


To compute the semantic similarity we resort on dynamic test executions since capturing all program behaviours is an undecidable problem. We thus, use a similarity coefficient, such as Ochiai coefficient, to compute the similarity of the passing and failing test cases. This is a common practice in many different lines of work such as mutation testing \cite{jia2009, JustJIEHF14, PapadakisSYB18, khanfir2020ibir} program repair \cite{GouesNFW12} and code analysis \cite{GoldBHIKY17} studies. Since semantic similarity compares the behavior between two program versions using a reference test suite, Ochiai coefficient~\cite{Ochiai} approximates program semantics using passing and failing test cases. 
Precisely, let $P_1$, $P_2$, $fTS_1$ and $fTS_2$ be two programs and their respectively set of failing tests, then the Ochiai coefficient between programs $P_1$ and $P_2$ is computed as $Ochiai(P_1,P_2) = \frac{|fTS_1 \cap fTS_2|}{\sqrt{|fTS_1| . |fTS_2|}}$, where $|\cdot|$ denotes the set size.


%% file: sections/motivating_example.tex
\section{Motivating example}
\label{sec:motivating-example}


We demonstrate the potential differences between syntactic and semantic deviations in fault seeding by using an example from the work of Tufano et al.~\cite{tufano2019learning}. Consider the following example\footnote{this example was taken from \cite[Figure~2]{tufano2019learning} and demonstrates a successful case where the fault seeded  by Tufano et al. matches exactly a real fault.}:

\begin{lstlisting}
//Original (abstracted) code in abstract representation (representation used by Tufano et al.)
public TYPE_1 remove ( int index ) { 
  TYPE_2 < TYPE_1 > VAR_1 = this . VAR_2 . remove ( index ) ; 
  return null != VAR_1 ? VAR_1 . get ( ) : null ; }
\end{lstlisting}

In this example, the \texttt{remove}  method first accesses one of the attributes of the invoking object (\texttt{this.VAR\_2}) and invokes recursively the method \texttt{remove}, saving the result in variable \texttt{VAR\_1}. Then, it returns \texttt{null} in the case that \texttt{VAR\_1} was \texttt{null}, otherwise it returns the result of invoking \texttt{VAR\_1.get()}. Tufano et al. seeds a fault that resembles exactly the real faulty instance, which is the following: 
\begin{lstlisting}
//Successful fault seeding with Tufano et al. (the fault resembles exactly the real fault)
public TYPE_1 remove ( int index ) { 
  return this . VAR_2 . remove ( index ) . get ( ) ; } 
\end{lstlisting}

The fault is caused because of the conditional check that is skipped and is indeed resembling a real fault made by developers~\cite{tufano2019learning}. In  particular the fault regards the checking of whether the result of the recursive call returns \texttt{null} or not that is ignored. 

Consider now a particular fault seeded by ``traditional'' mutation testing, using simple syntactic changes (e.g., generated by the REMOVE\_CONDITIONALS \footnote{https://pitest.org/quickstart/mutators/\#REMOVE\_CONDITIONALS} operator from PiTest~\cite{ColesLHPV16}): 
\begin{lstlisting}
//Fault seeded using mutation testing, simple syntax-based mutation
public TYPE_1 remove ( int index ) { 
  TYPE_2 < TYPE_1 > VAR_1 = this . VAR_2 . remove ( index ) ; 
  return true ? VAR_1 . get ( ) : null ; } 
\end{lstlisting}

This mutant replaces the condition \texttt{null != VAR\_1} by \texttt{true}, making the guarded statements (i.e. \texttt{VAR\_1.get()}) to be executed irrespective of the condition (actually it is always executed). 

Interestingly, by comparing the two faulty instances one can easily observe that  they are syntactically different despite being semantically equivalent. One can also observe that a simple syntactic transformation, such as the one used by mutation testing perfectly matches the complex transformation learned by Tufano et al. To make the differences concrete we can compute the BLEU scores (syntactic similarity  between seeded and real fault), i.e., the evaluation metric used by Tufano et al.~\cite{tufano2019learning}, and see that the returned scores are \emph{1} and \emph{0.48}, respectivelly. However, as the mutants are equivalent and resemble a real fault, their semantic similarity is 1 despite the large difference in the BLEU scores. 

The above example, clearly evidence that seeded faults do not necessarily need to be similar to real faults in order to resemble them. At the same time, the above example demonstrates the \textit{fault coupling}~\cite{offut1992}, i.e., simple syntactic transformations, such as those used by mutation testing, couple to more complex faults. In this particular case, the transformation performed by mutation is significantly smaller than Tufano et al. as it has a BLEU score (syntactic similarity from the original code) 0.85 while, Tufano et al. has 0.39.

%% file: sections/research_questions.tex
\section{Research questions}
\label{sec:research-questions}
We start our analysis by recording the syntactic and semantic similarity between injected and real faults. We perform this analysis 
to understand the general relation between seeded and real faults and check if there is any associations between these two variables. The existence of such a relation will provide evidence that fault seeding techniques, instead of using grammar-based (simple) transformations as it is traditionally done in mutation testing, should attempt to form frequent fault patterns and design fault seeding techniques guided by real fault instances, in a sense follow similar path to static code analysis \cite{PradelS18, HovemeyerP04}. Therefore, we ask: 

\begin{description}
    \item[RQ1] \emph{How semantically and syntactically similar are seeded and real faults?}
\end{description}

The answer to this question will also provide evidence on the use of syntactic distance metrics in evaluating fault seeding methods, i.e., whether seeded faults with small syntactic distance from the real ones are indeed semantically close to them (at least closer than those that are not syntactically similar, i.e., dissimilar). As we discussed before, such evaluation metrics are followed by recent research (Table \ref{tbl:Related}) without any empirical evidence of their validity. This means that we want to check whether the choice of syntactic distance as an evaluation metric, as performed by previous studies in Table \ref{tbl:Related}, is a valid choice.

To further investigate the fault-revealing potential of the techniques, we also check whether real faults can be resembled solely by syntactically similar (but not exact matches) or dissimilar seeded faults. The difference of this analysis with the above is that instead of interest in a general relationship, here we are interested in the existence of some, perhaps few, cases that resemble the sought faults. In practical terms, the existence of such instances indicates the minimum level of fault detection that can be achieved by fault-based testing when one attempts to design test cases that reveal all seeded faults \cite{JustJIEHF14}. 
Thus, we ask:

\begin{description}
    \item[RQ2] \emph{How many real faults we can resemble by using syntactically similar and dissimilar seeded faults?}
\end{description}




Having investigated the relationships between syntactic changes and semantic changes, we turn our attention to comparing the fault seeding techniques. Precisely, we seek to examine whether there is any utility benefit in combining the techniques. Therefore, we ask:

\begin{description}
    \item[RQ3] \emph{How  do the  employed techniques compare to each other? Are they producing faults subsumed by mutation testing?}
\end{description}

This question aims to study the ``overlaps'' of the fault seeding techniques to strengthen fault seeding tools. We investigate the practical use of the approaches when testers aim at revealing the seeded faults and compare the techniques in a cost-benefit fashion. 

%% file: sections/tools.tex
\section{Fault Seeding}
\label{mutation_testing_tools}

We employ the following fault seeding techniques: 
\begin{itemize}[leftmargin=5.5mm]
    \item PiTest~\cite{ColesLHPV16} a mutation testing technique that uses simple grammar-based transformations for seeding faults. 
    \item IBIR~\cite{khanfir2020ibir} a fault mimicking technique that attempts to seed faults using manually crafted fault patterns (based on fault analysis of common fault instances). 
    \item DeepMutation~\cite{DeepMutation} a fault mimicking technique that learns to seed faults based on bug-fixes, i.e., it learns inverted bug-fixes. 
    \item CodeBERT, a mutation testing technique that we develop. CodeBERT seeds ``natural'' faults by replacing tokens based on code embeddings~\cite{DBLP:conf/emnlp/FengGTDFGS0LJZ20}.
    CodeBERT seeds faults that follow the implicit programming norms of programs captured by big code. 
\end{itemize}

\subsection{PiTest}

PiTest~\cite{ColesLHPV16} is a state-of-the-art mutation testing tool for Java. It seeds faults by manipulating the bytecode of programs in order to 
avoid mutant compilation and reduce mutant generation overheads. 
PiTest works by analyzing entire bytecode sequences and by looking for a possible location, i.e., instruction, to seed faults. 
Mutants Operators are designed to be stable and categorized into 29 task-specific distinct groups. Hence, an example of groups is Conditionals Boundary and Return Values mutators, which seed variations concerning relational operators and method call return values. PiTest provides over 120 Mutant Operators, among which are many experimental mutants used for scientific purposes. For this study, we take into consideration ALL mutants that PiTest offers. 
Every mutant is uniquely identified, providing sufficient information to replicate the bytecode transformation to compute syntactic and semantic similarity scores. We employ ASM Java bytecode instrumentation API for this purpose.


\subsection{IBIR}
IBIR~\cite{khanfir2020ibir} is a Java fault seeding tool that leverages an information-retrieval-based fault localization model (IRFL) combined with automatic program repair (APR) inverted fix-patterns. It aims to favour the generation of few but realistic mutants (similar to real ones). It takes as input the git repository of the program to mutate and a bug report of this latter, written in natural language. It then seeds multiple faults (inducing multiple faulty versions of the program) that emulates the fault described in the inputted bug report. 

IBIR starts by analysing the given bug report using IRFL ~\cite{ZhouZL12} to identify locations that are likely to be related to the features impacted by the corresponding fault.
Second, IBIR applies fault-patterns on the identified locations, which are inverted fix-patterns used in pattern-based automated program repair approaches~\cite{khanfir2020ibir}. As the fix-patterns are crafted from real bug-fixes, their inverse would induce faults similar to real faults. 
IBIR repeats this process until all location-pattern pairs have been treated, or a predefined number of faults is reached, which is set to 100 by default. In this study we passed to its provided filtering parameter, the classes changed by the bug-fix on defects4j to exclude the mutants from other classes, and we run it with the rest of its setup, as default.

\subsection{DeepMutation}
\emph{DeepMutation}~\cite{DeepMutation} generates mutants using a Neural Machine Translation (NMT) \cite{tufano2019learning} trained on a large corpus ($\sim$787k) of existing bug-fixing commits mined from GitHub repositories. 
It takes as input a Java class file and starts by extracting the methods to mutate. 
Then, it applies an abstraction process to each method (i.e., user-defined names are replaced by learnable identifiers) to obtain an abstracted representation for each method (like the one shown in Section~\ref{sec:motivating-example}). These are then fed into the pre-trained NMT model to produce abstract mutations translated back to source code. 

We followed the guidelines provided by the DeepMutation tool~\cite{deepmutationweb},  but unfortunately we could not install and run the provided implementation of DeepMutation. We thus, reproduced it by using the publicly available pre-trained model and \texttt{src2abs}~\cite{src2abs} tool to abstract and translate back to source code the mutants. We generate one mutant per method as done by DeepMutation~\cite{DeepMutation}.

\subsection{CodeBERT based Mutation}
\emph{CodeBERT}~\cite{codebertweb,DBLP:conf/emnlp/FengGTDFGS0LJZ20} is a pre-trained language model for programming languages, recently released by Microsoft Research. 
It was pre-trained on around 6.4 million programs from 6 different programming languages, including Java. It supports a variety of tasks, such as natural language code search and code documentation generation, but we are only using the Masked Language Modeling (MLM) task. MLM takes as input a text sequence from which there is 1 masked token, which it attempts to predict (replace). For instance, given the masked sequence \texttt{int a = <mask>;}, CodeBERT predicts that \texttt{0}, \texttt{1}, \texttt{b}, \texttt{2},  and \texttt{10} are the (five) most likely tokens to replace the masked one (ordered in descending order according to their score -- likelihood). 

We use CodeBERT to generate mutants by selecting tokens to mask and replacing them with the predicted tokens. Specifically, we develop a mutant generation tool, based on CodeBERT's predictions, that iterates on each statement and repeats the following process: (1) it selects and masks one token at a time, depending on the type of expression being analysed; (2) it feeds CodeBERT with the masked sequence and obtains the predictions; (3) it creates mutants by replacing the masked token with the predicted ones; and (4) it discards non-compilable and duplicate mutants (mutants syntactically the equal to original code). 
CodeBERT uses multi-layer bidirectional Transformer~\cite{VaswaniSPULALP2017}, meaning that the predictions are context-dependent, which result in different mutants when masking the same variable in different program locations. 

Our implementation applies to a wide variety of Java expressions, being able to mutate operands and operators of unary/binary expressions, assignment statements, literals, variable names, method calls, object field accesses, among others. This indicates that for the same program location, several mutants can be generated. For instance, for a binary expression like \texttt{a == b}, our tool will create (potentially 15) mutants from the following 3 masked sequences: \texttt{<mask> == b}, \texttt{a <mask> b}, and \texttt{a == <mask>}. 
We feed a sequence of 512 tokens to CodeBERT (maximum sequence length supported) that includes the masked token and its context.

%% file: sections/experimental_setup.tex
\section{Experimental setup}
\label{sec:experimetal-setup}

\subsection{Real Faults}

 We used Defects4J 
 \cite{defects4j} v2.0,0, which contains over 800 faults with supporting build infrastructure and forms one of the largest collections of reproducible real faults for Java programs.
 
 Every fault in the dataset consists of the faulty and fixed versions of the code, a developer's test suite accompanying the project, and information regarding the commit modified classes and the patches produced to fix the fault. The faults have manually been minimized, so every irrelevant change to the fix has been removed. The dataset also includes at least one fault triggering test that fails in the faulty version and passes in the fixed one.  

The set of faults spans more than a decade of development history, making it challenging for us to synchronize the execution of faults over different mutant generation techniques. 
Thus, intending to be as fair as possible with the selected tools, we had to discard those faults that do not satisfy the building requirements for each tool. Precisely, we discarded the faults from the project Jfreechart (number of faults 26) and Closure-compiler (174). Additionally, at the time of conducting this study, we found that 82 faults from Jsoup project were not compilable due to technical reasons~\cite{defects4j_issue}.

We consider the following projects and number of faults from Apache Commons~\cite{apache_commons} 
family, which is a collection of projects of Java utility classes: commons-cli (38), commons-codec (18), commons-compress (33), commons-csv (16), commons-math (100), commons-lang (63), commons-collections (4), commons-jxpath (21). We also include projects from the Jackson~\cite{jackson} 
family, which is a suite of data-processing tools for Java and includes jackson-core (26), jackson-databind (102), and jackson-dataformat-xml (6). 
Additionaly, we include faults from Mockito (27), one of the most popular mocking frameworks in Java, Jsoup (11), a Java library for HTML parsing,
Gson (18) a Java library for JSON parsing and generation from and into java objects,
and joda-time (26) for the Java date and time classes. In total, we analyzed 509 faults from 15 different projects.

\subsection{Seeded Faults} 
For each selected faulty project version from Defects4J, we start by identifying the modified classes between the faulty and fixed versions. Then we generate mutants for the fixed version of each modified class by employing the selected mutation testing tools. 

Table \ref{tbl:tools_mutants_statistics} records the number of faults analysed and the number of mutants generated by each mutation testing tool. 
Since we use all available mutation operators provided by PiTest, it generated 1,120,719 mutants for the 509 faulty classes.
Our CodeBERT-based mutation testing technique applied on 499 faults and produced a total of 192,905 mutants. DeepMutation, which produces only 1 mutant per method, produced 3,057 mutants for the 348 analysed faults. 
IBIR produces at most 100 number of mutants per bug report, resulting in a total of 38,327 mutants for the 393 analysed faults.

\begin{table}[tp]
\vspace{-1.0em}
\centering
\caption{Mutants used}
\vspace{-1.2em}
\resizebox{\columnwidth}{!}{
\scriptsize{
\begin{tabular}{l|r|r}
\toprule
\textbf{\scriptsize{Fault Seeding Tool}} & \textbf{\scriptsize{\# of Analysed Faults}} & \textbf{\scriptsize{\# of Mutants}} \\
\midrule
\scriptsize{PiTest} & \scriptsize{509} & \scriptsize{1,120,719} \\
\scriptsize{CodeBERT} & \scriptsize{499} & 192,905 \\
\scriptsize{DeepMutation} & 348 &  3,057\\
\scriptsize{IBIR} & \scriptsize{393}  & \scriptsize{38,327}  \\
\bottomrule
\end{tabular}
}
}
\label{tbl:tools_mutants_statistics}
\vspace{-1.6em}
\end{table}

\subsection{Experimental Procedure}

We start by executing every mutant generated using the Defects4J framework, recording the set of failing tests (if any) killing each mutant. 
Then we proceed to compute syntactic and semantic similarities between the mutants and the corresponding faults, relying on the metrics defined in Section~\ref{mutation_testing}. Thus, syntactic similarity between the mutant and the fault will be measured in terms of the BLEU score, while the semantic similarity will be characterised by the Ochiai coefficient between the mutant and the fault. 
It is worth to mention that, since PiTest produces the mutations at bytecode level, we perform the syntactic similarity computation between the bytecode sequences corresponding to mutants and faults. 

To answer RQ1, we study the existence or not of any correlation between the syntactic and semantic similarity between seeded and real faults. 
We do this analysis per tool bases, and firstly by considering all the mutants created with each particular tool (all mutants), and secondly by considering only the mutants that change the same method that was modified in the patch from  fault to fix  according to the information gathered from Defects4J (modified-methods mutants). 
In the first case, by analysing all the mutants, we are trying to see if some general trend is observed, even when seeded faults are changing locations that are not part of patches (leading to lower syntactic similarities). 
In the second case we aim at controlling this, by analysing only the mutants from the same methods modified in the patches (leading to higher syntactic similarities). 

We then study whether high scores for syntactic similarity (i.e., seeded and real faults are syntactically similar) implies high scores for semantic similarity (i.e., seeded and real faults behave the same). And dually, we also explore whether low scores for syntactic metrics implies low scores for semantic metrics. 
To do so, for each fault and tool, we start by sorting mutants in ascending order according to their syntactic similarity, and we organise them into four quartiles $Q_1$, $Q_2$, $Q_3$ and $Q_4$, where $Q_1$ represents the most syntactically dissimilar mutants wrt the fault (lowest syntactic scores) and $Q_4$ represents the most syntactically similar mutants wrt the fault (highest syntactic scores). 
Finally, for the mutants in each quartile we analyse their semantic similarity wrt the fault, with the aim at observing if there is any evidence that more syntactically  dissimilar mutants behave very differently than the faults, and whether syntactically similar mutants behave the same as the faults. 

To answer RQ2, we study if there is at least one mutant that behaves the same as the fault. Since checking if two programs are equivalent is undecidable, we will consider that a mutant resemble a real fault whether they behave the same wrt the test suite, i.e., when the semantic similarity (Ochiai) value is equals to 1. 
Firstly, we start by analysing the percentage of real faults that each tool can resemble (i.e., percentage of real faults for which the tool generated at least one mutant with Ochiai 1).  
Secondly, we focus the analysis on the same quartiles used for RQ1 to study the percentage of real faults resembled by syntactically similar and dissimilar mutants. 


Up to this point our analysis checks single mutant instances.  However, mutation testing tools are generating sets of mutants, which may have different properties than the single instances \cite{PapadakisCT18}.
Thus, to answer RQ3, we compare the mutants generated with every technique with respect to following aspects:
\begin{itemize}[leftmargin=8.0mm]
    \item We make an objective comparison between the techniques \cite{LaurentPKHTV17} in terms of mutant kills and faults detected. In a sense we select minimal test cases, from the developer test suites, that kill the maximum number of mutants of each technique and check how well they kill mutants from the other techniques and whether they detect the associated real faults or not. This analysis checks how well the test requirements (mutants) of one technique work against the test requirements of the other. 
    \item We study  the subsumptions~\cite{KurtzADOD2014} between mutants produced by the different tools (mutant $m_1$ subsumes mutant $m_2$, if every test that kills $m_1$, also kills $m_2$~\cite{KurtzADOD2014}) with respect to their joint use. Subsuming mutants have been proposed as a way to evaluate mutant selection methods \cite{ammann_establishing_2014, PapadakisHHJT16}. In essence they quantify the contributions of mutant sets  by discarding subsumed mutants \cite{PapadakisK00TH19}. We thus compute the subsumption relations among all mutants produced by all techniques and count the number of mutants that are in the top of the hierarchy and are not subsumed by each tool. This cross-technique evaluation indicates the overlaps between the techniques and highlights opportunities for complementary use of the techniques. 
    \item We simulate a testing scenario where a tester selects a subset of mutants, to use for mutation analysis, and designs tests to kill them. We start by taking each mutant created by one tool, selecting a test that kills it, until we reach a maximum of mutants killed. The selected test suite is then used to measure mutation score from the pool of mutants created by the other tools. We repeat this process 100 times to reduce the impact of the random selection of killing tests on our results. This analysis form a cost-effective evaluation and aims at emphasising the effects of the different approaches  \cite{KurtzAODKG16, AndrewsBLN06}. Here it must be noted that  in order to make the comparison fair, in RQ3 we investigate only the classes-faults analysed by all the tools. 
\end{itemize}



\subsection{Statistical Analysis}

To study the correlation between semantic and syntactic properties, we use two correlation metrics: Kendall rank coefficient ($\tau$)) (Tau-a) and Pearson product-moment correlation coefficient (r). In every case, we take into consideration the 0.05 significance level. Each correlation coefficients measure similarity, taking values from -1 to 1. Values close to both ends represent negative and positive correlation, respectively. While values in a range of absolute 0.2 around zero, donate absence and insignificant correlation.

%% file: sections/empirical_evaluation.tex
\section{Empirical Evaluation}

\subsection{Syntactic and Semantic Similarity between Seeded and Real Faults (RQ1)}

\begin{figure*}[!htb]
\centering
\vspace{-0.5em}
\begin{subfigure}[b]{0.23\textwidth}
\centering
\includegraphics[width=\textwidth]{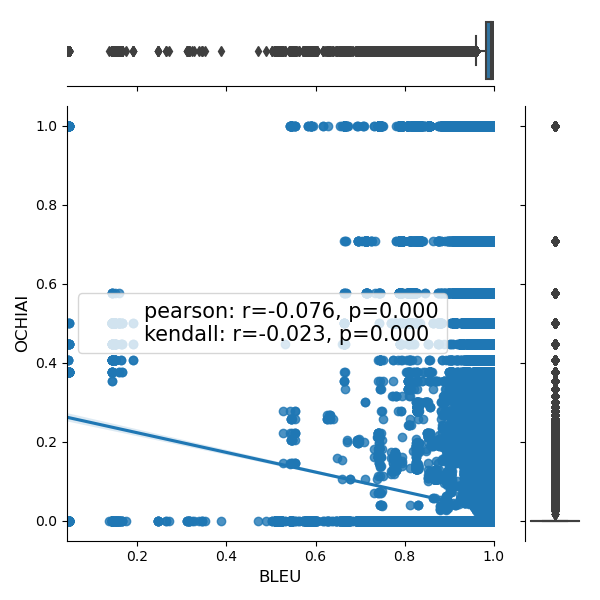}
\vspace{-1.2em}
\caption[PiTest]%
{{\small PiTest}}    
\label{fig:PitTest_class_level_rq1}
\end{subfigure}
\begin{subfigure}[b]{0.23\textwidth}
\centering
    \includegraphics[width=\textwidth]{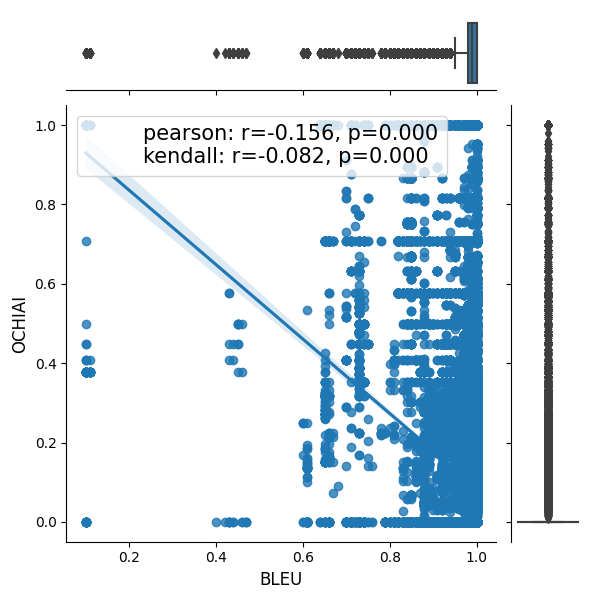}
\vspace{-1.2em}
\caption[CodeBERT]%
{{\small CodeBERT}}
\label{fig:Codebert_class_level_rq1}
\end{subfigure}
\begin{subfigure}[b]{0.23\textwidth}
\centering
    \includegraphics[width=\textwidth]{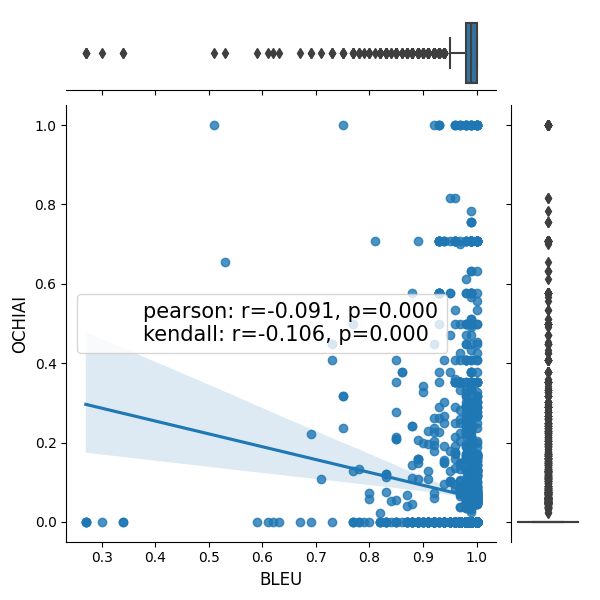}
\vspace{-1.2em}
\caption[DeepMutation]%
{{\small DeepMutation}}    
\label{fig:DeepMutation_class_level_rq1}
\end{subfigure}
\begin{subfigure}[b]{0.23\textwidth}
\centering
    \includegraphics[width=\textwidth]{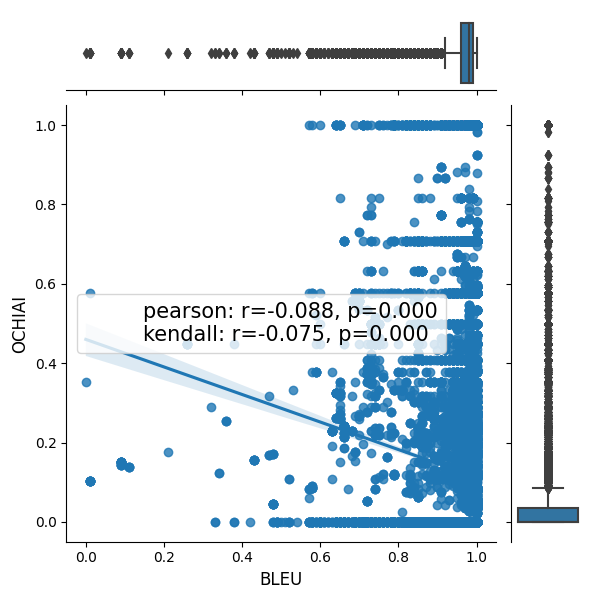}
\vspace{-1.2em}
\caption[IBIR]%
{{\small IBIR}}    
\label{fig:IBIR_class_level_rq1}
\end{subfigure}
\vspace{-1.0em}
\caption{\small RQ1: Syntactic vs semantic similarity at class granularity level. There's a weak link between syntactic and semantic similarity.}
\label{fig:RQ1-all-mutants}
\end{figure*}

Figure~\ref{fig:RQ1-all-mutants} summarises the syntactic and semantic similarity values between the mutants created with different tools and the real faults. Interestingly, we notice that while many of the mutants have high syntactic similarity, their semantic similarity is scattered from 0 to 1. This seems to imply that the relationship between the two metrics is weak.

\begin{figure*}[!htb]
\vspace{-1.0em}
\centering
\begin{subfigure}[b]{0.23\textwidth}
\centering
    \includegraphics[width=\textwidth]{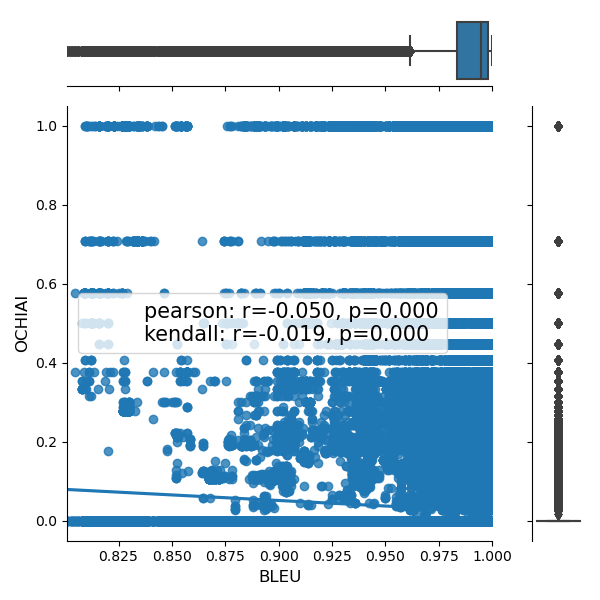}
\vspace{-1.2em}
\caption[PiTest]%
{{\small PiTest}}    
\label{fig:PitTest_function_level_rq1}
\end{subfigure}
\begin{subfigure}[b]{0.23\textwidth}
\centering
    \includegraphics[width=\textwidth]{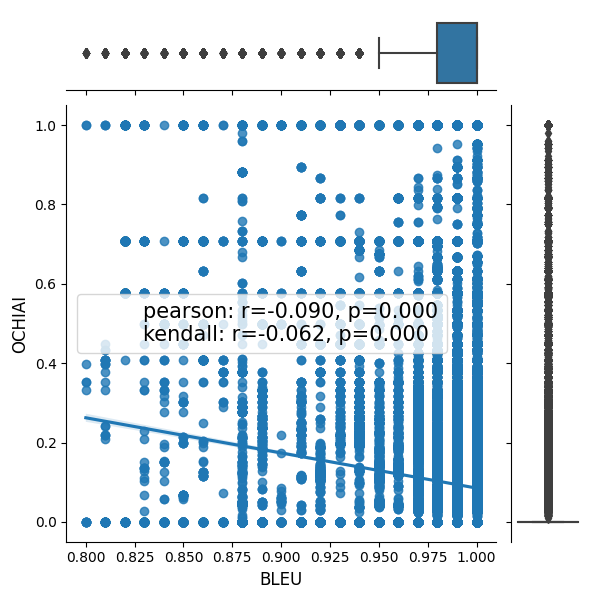}
\vspace{-1.2em}
\caption[CodeBERT]%
{{\small CodeBERT}}    
\label{fig:Codebert_function_level_rq1}
\end{subfigure}
\begin{subfigure}[b]{0.23\textwidth}
\centering
    \includegraphics[width=\textwidth]{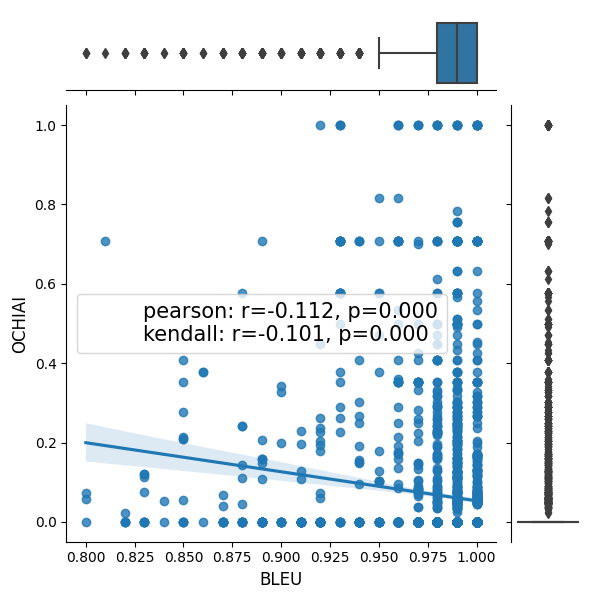}
\vspace{-1.2em}
\caption[DeepMutation]%
{{\small DeepMutation}}    
\label{fig:DeepMutation_function_level_rq1}
\end{subfigure}
\begin{subfigure}[b]{0.23\textwidth}
\centering
    \includegraphics[width=\textwidth]{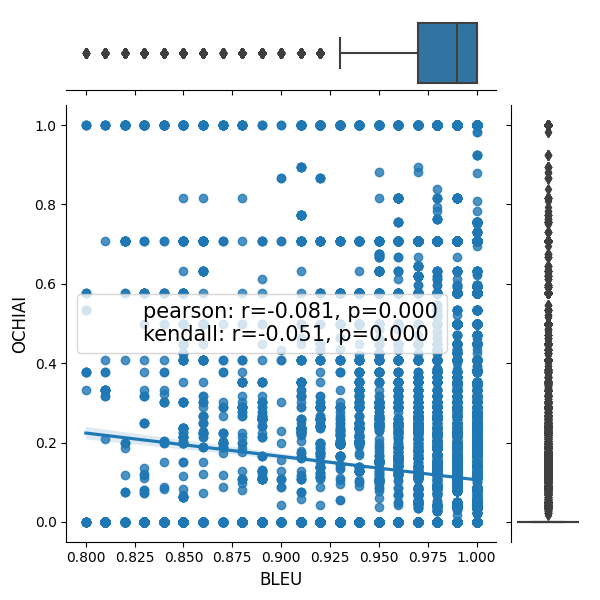}
\vspace{-1.2em}
\caption[IBIR]%
{{\small IBIR}}    
\label{fig:IBIR_function_level_rq1}
\end{subfigure}
\vspace{-1.0em}
\caption{\small RQ1: Semantic vs syntactic similarity when syntactic similarity is greater than 80\%. Again no link between them.}
\label{fig:RQ1-syntactic-80}
\end{figure*}

To check this further, in Figure~\ref{fig:RQ1-syntactic-80} we depict syntactic and semantic similarity values for all mutants with a syntactic similarity greater than 0.8 (i.e., mutants syntactically very similar to real faults). We notice that the four tools generated mutants that behave as faults (obtaining Ochiai 1). These are both syntactically similar and dissimilar to the faults (see the plots' top values, y-axis). We can also observe that most of the mutants that are syntactically almost the same as real faults (BLEU near 1) behave very differently (Ochiai near 0), indicating that the relationship is weak even when seeded faults are close to real ones.

\begin{figure*}[!htb]
\centering
\vspace{-1.0em}
\begin{subfigure}[b]{0.23\textwidth}
\centering
    \includegraphics[width=\textwidth]{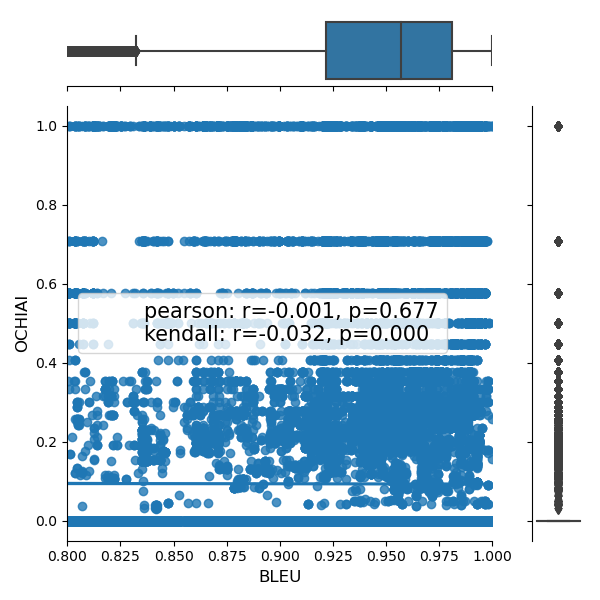}
\vspace{-1.2em}
\caption[PiTest]%
{{\small PiTest}}    
\label{fig:PitTest_function_level_rq1}
\end{subfigure}
\begin{subfigure}[b]{0.23\textwidth}
\centering
\includegraphics[width=\textwidth]{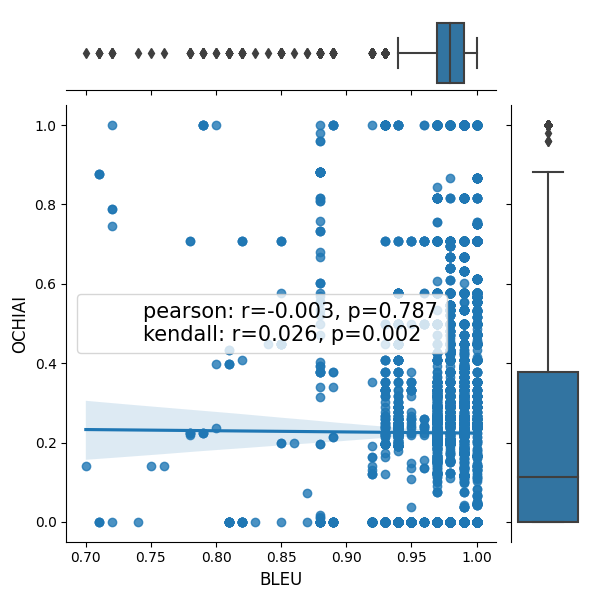}
\vspace{-1.2em}
\caption[CodeBERT]%
{{\small CodeBERT}}    
\label{fig:Codebert_function_level_rq1}
\end{subfigure}
\begin{subfigure}[b]{0.23\textwidth}
\centering
    \includegraphics[width=\textwidth]{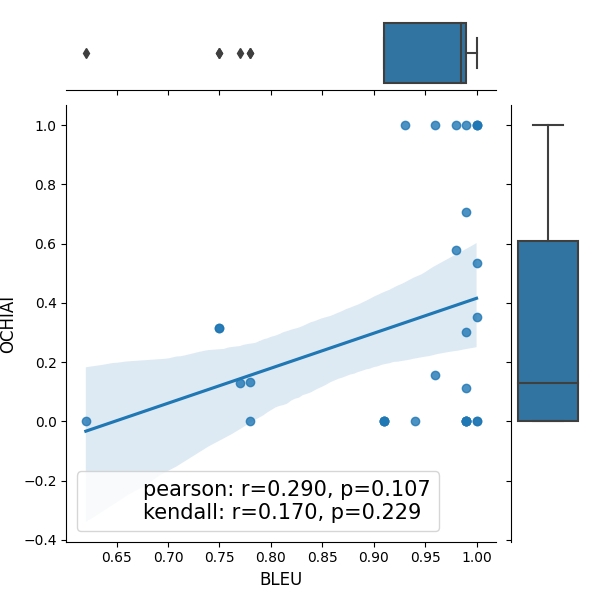}
\vspace{-1.2em}
\caption[DeepMutation]%
{{\small DeepMutation}}    
\label{fig:DeepMutation_function_level_rq1}
\end{subfigure}
\begin{subfigure}[b]{0.23\textwidth}
\centering
    \includegraphics[width=\textwidth]{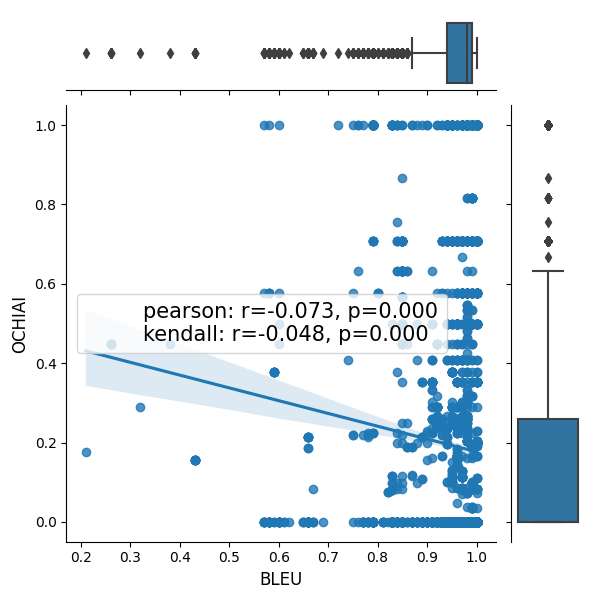}
\vspace{-1.2em}
\caption[IBIR]%
{{\small IBIR}}    
\label{fig:IBIR_function_level_rq1}
\end{subfigure}
\vspace{-1.0em}
\caption[Correlation between semantic and syntactic similarity of seeded and real faults]{\small RQ1: Semantic vs syntactic similarity at method granularity. Faults from the same method have divers semantic similarity.}
\label{fig:RQ1-modified-methods}
\end{figure*}

The above results are based on the class granularity level, and therefore their syntactic and semantic changes may be impacted by the ``size'' of the seeded faults. We thus, analyse the results at method level granularity as well. Figure~\ref{fig:RQ1-modified-methods} shows the syntactic and semantic similarities only for the mutants that change the same methods as the real faults. 
In this case, we focus on mutants with syntactic similarity greater than 80\% and see a similar trend with the class level results,  i.e., every tool managed to generate syntactically similar and dissimilar mutants that behave exactly like a real fault. Additionally, the syntactic similarity is almost 1, but the semantic similarity is scattered from 0 to 1, as in the class level granularity case.
Therefore, it indicates a potential threat to validity to the studies that approximate semantic similarity through the use of a syntactic one. 

To further analyze this relationship and look more into its practical use case and investigate whether seeding faults with small syntactic distance from the real ones result in semantically close faults to real ones (at least closer than those that are not syntactically similar, i.e., dissimilar).  The key objective is to check whether there is some effect when we have high syntactic similarity.  

\begin{figure*}[!htb]
\vspace{-1.2em}
\centering
\begin{subfigure}[b]{0.23\textwidth}
    \centering
    \includegraphics[width=\textwidth]{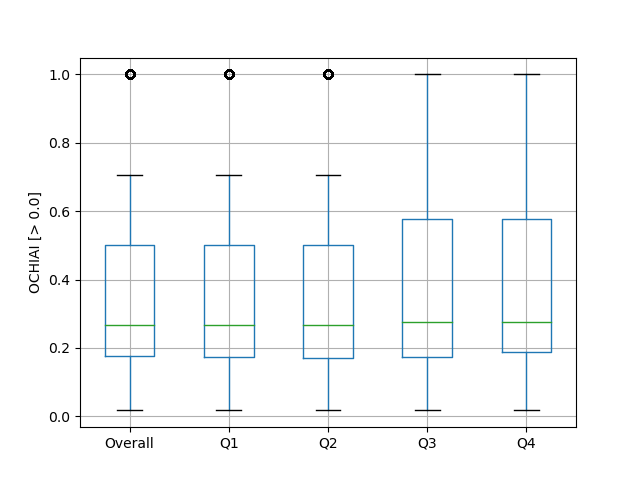}
    \vspace{-1.2em}
    \caption[PiTest]%
    {{\small PiTest}}    
    \label{fig:PiTest_rq2}
\end{subfigure}
\begin{subfigure}[b]{0.23\textwidth}
    \centering
    \includegraphics[width=\textwidth]{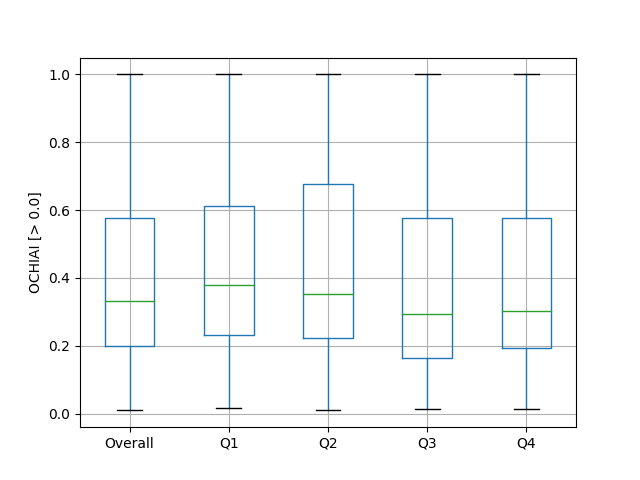}
    \vspace{-1.2em}
    \caption[CodeBERT]%
    {{\small CodeBERT}}    
    \label{fig:Codebert_rq2}
\end{subfigure}
\begin{subfigure}[b]{0.23\textwidth}
\centering
    \includegraphics[width=\textwidth]{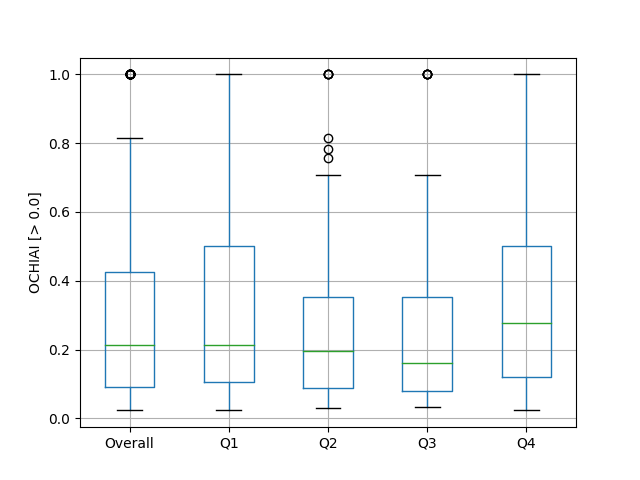}
    \vspace{-1.2em}
\caption[DeepMutation]%
{{\small DeepMutation}}    
\label{fig:DeepMutation_rq2}
\end{subfigure}
\begin{subfigure}[b]{0.23\textwidth}
        \centering
        \includegraphics[width=\textwidth]{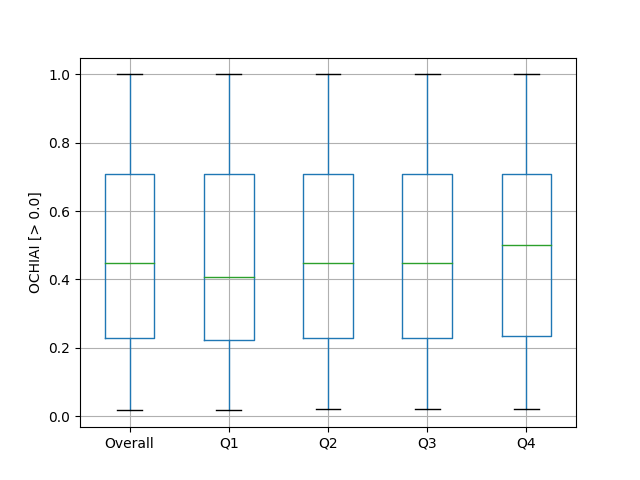}
        \vspace{-1.2em}
        \caption[IBIR]%
        {{\small IBIR}}    
        \label{fig:PiTest_rq2_mutants_from_random_lines}
\end{subfigure}
\vspace{-1.0em}
\caption{\small RQ1: Distribution of Semantic similarities across different levels of syntactic similarities.}
\label{fig:RQ2-syntactic-proxy-for-semantic}
\vspace{-1.0em}
\end{figure*}

Figure~\ref{fig:RQ2-syntactic-proxy-for-semantic} shows the distribution of semantic similarities when we group mutants according to their syntactic similarity. We observe that semantic similarity is uniformly distributed between mutants that are syntactically similar and dissimilar to the real faults.  This evidence that more minor syntactic transformations do not imply smaller semantic changes, and at the same time, more significant syntactic changes also do not imply more extensive semantic changes. \\



\begin{tcolorbox}[
    standard jigsaw,
    opacityback=0,
    left=0pt,right=0pt,top=0pt,bottom=0pt, arc=0pt,
    boxrule=0.4pt,leftrule=1.5pt
]
Many seeded faults behave quite similarly to real faults (high semantic similarity) while at the same time having low syntactical similarity to real faults. Perhaps surprisingly, we find no evidence suggesting any link between syntactic and semantic similarity, except from the cases of exact matches.
\end{tcolorbox}

\subsection{Mutants Resembling Real Faults (RQ2)}
Table~\ref{tbl:rq3a} summarizes the percentage of real faults that were resembled by at least one mutant produced by each tool.  Column \emph{Total} indicates the percentage of real faults for which the tool generated at least one mutant with semantic similarity 1, meaning that it semantically resembles the real fault. Column \emph{Exact Matches} indicates the percentage of real faults for which the tools produced one mutant syntactically the same as the fault. Columns \emph{Q1, Q2, Q3} and \emph{Q4} express the percentage of faults semantically resembled by mutants of each quartile. Notice that quartile \emph{Q4} includes mutants syntactically the same as real faults, which have syntactic similarity 1 (this explains why Q4 results are more inflated than for the other quartiles). 

We observe that PiTest resembles 46.85\% of the real faults, while CodeBERT successfully resembles 60.42\%, DeepMutation 6.9\% and IBIR 43.01\%. Interestingly, CodeBERT reproduces most of the real faults, significantly more than the other tools:  13.57\% more than PiTest, 17.41\% more than IBIR and 53.52\% more than DeepMutation. A reason why PiTest performance is lower is that its mutants are designed to be hard to be detected by test suites, i.e., they aim at producing small semantic changes to programs, so the number of failing tests for them can be very small wrt the number of failing tests for the real faults (leading to lower Ochiai). In the case of IBIR, the reason is the opposite; it tends to produce several transformations to resemble real faults, that may lead to big semantic differences.  In the case of DeepMutation, it manages to reproduce real faults even when we are creating only one mutant per method but these are very few cases. 
Regarding exact matching, IBIR successfully reproduces 5.6\% of the faults, outperforming the rest of the tools which only managed to reproduce less than 1\%. 

\begin{table}[tp]
\centering
\vspace{-0.7em}
\caption{RQ2 Percentage of real faults with at least one semantically similar mutant - Quartiles represent mutants sorted by syntactic similarity}
\vspace{-0.95em}
\resizebox{\columnwidth}{!}{
{\small
\begin{tabular}{l|r|r|r|r|r|r}
\toprule
\textbf{\scriptsize{Fault Seeding Tool}} & \textbf{\scriptsize{Total}} & \textbf{\scriptsize{Exact Matches
}} &  \textbf{\scriptsize{Q1}} & \textbf{\scriptsize{Q2}} & \textbf{\scriptsize{Q3}} & \textbf{\scriptsize{Q4}}  \\
\midrule
PiTest          & 46.85 & 0.58 & 29.20 & 29.61 & 32.45 & 39.09 \\
CodeBERT        & 60.42 & 0.42 & 23.73 & 25.68 & 27.91 & 28.42 \\
DeepMutation    & 6.90  & 0.57 & 1.84 & 1.34  & 1.95 & 4.31\\
IBIR            & 43.01 & 5.60 & 26.21 & 27.48 & 28.24 & 33.84 \\
\bottomrule
\end{tabular}
}
}
\label{tbl:rq3a}
\vspace{-0.8em}
\end{table}

Table~\ref{tbl:rq3b} records the mean ratios of mutants that behave as the real faults. 
We observe that in average between 3.55\%-4.9\% of mutants resemble the real faults,  independently of their syntactic similarity. \\

\begin{table}[tp]
\centering
\vspace{-0.5em}
\caption{RQ2: Mean ratios of mutants resemble real faults.}
\vspace{-0.95em}
\resizebox{\columnwidth}{!}{
{\scriptsize
\begin{tabular}{l|r|r|r|r|r}
\toprule
\textbf{\scriptsize{Fault Seeding Tool}} & \textbf{\scriptsize{Overall}} &  \textbf{\scriptsize{Q1}} & \textbf{\scriptsize{Q2}} & \textbf{\scriptsize{Q3}} & \textbf{\scriptsize{Q4}}  \\
\midrule
PiTest          & 4.10 & 0.85 & 1.00 & 1.07 & 1.19 \\
CodeBERT        & 4.30 & 4.33 & 4.18 & 4.17 & 4.50 \\
DeepMutation    & 3.55 & 1.46 &	1.12 & 1.45 & 3.85 \\
IBIR            & 4.90 & 4.72 &	4.31 & 5.03 & 5.54 \\
\bottomrule
\end{tabular}
}
}
\label{tbl:rq3b}
\vspace{-1.3em}
\end{table}

\begin{tcolorbox}[
    standard jigsaw,
    opacityback=0,
    left=0pt,right=0pt,top=0pt,bottom=0pt, arc=0pt,
    boxrule=0.4pt,leftrule=1.5pt
]
CodeBERT resembles 60.42\% of the real faults, significantly outperforming PiTest (46.85\%), IBIR (43.01\%) and DeepMutation (6.9\%). 
In terms of exact (syntactic) matching, IBIR reproduces 5.6\% of the faults, outperforming the other tools. Interestingly, 3.55\%-4.9\% of the mutants behave as real faults. 
\end{tcolorbox}

\subsection{Comparing fault seeding techniques (RQ3)}
\begin{figure*}[!htb]
\vspace{-0.8em}
\centering
\begin{subfigure}[b]{0.325\textwidth}
\centering
\includegraphics[width=\textwidth]{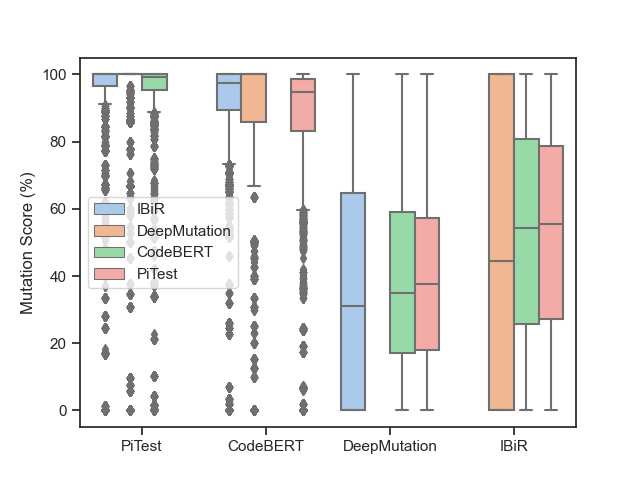}
\vspace{-1.4em}
\caption{{\small Mutation scores achieved on the mutants of the 4 techniques when selecting test cases that kill the mutants of each one. Higher scores indicate higher collateral kills among the techniques (semantic overlaps between the seeded faults).}}
\label{fig:RQ5-tools-simulation}
\end{subfigure}
\hspace{0.8mm}
\begin{subfigure}[b]{0.32\textwidth}
\centering
\includegraphics[width=\textwidth]{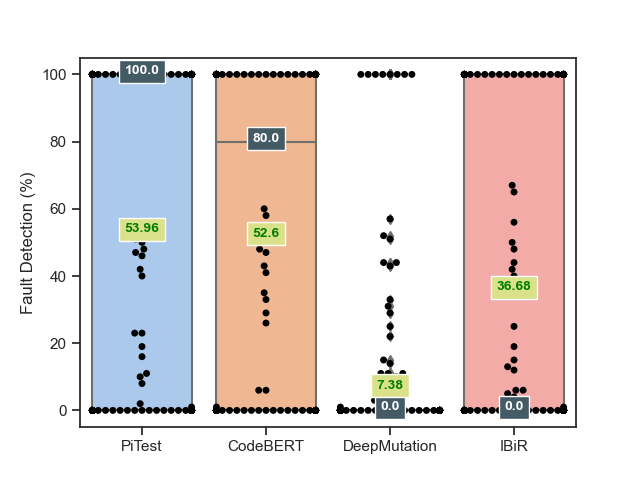}
\vspace{-1.4em}
\caption{{\small Fault detection of test cases that kill the mutants of the techniques. PiTest and CodeBERT perform best with median-average values of 100\%-53.96\% and 80\%-52.6\%. IBIR and DeepMutation achieve median-average values of 0\%-7.36\% and 0\%-36.68\%.}}
\label{fig:RQ5-revailing_bugs}
\end{subfigure}
\hspace{0.8mm}
\begin{subfigure}[b]{0.325\textwidth}
\centering
\includegraphics[width=\textwidth]{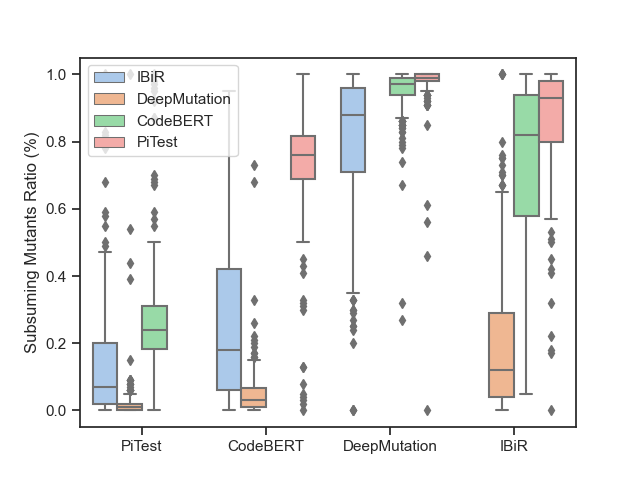}
\vspace{-1.4em}
\caption{{\small Technique's subsumption. Ratios of subsuming mutants contributed by the joint use of one technique with the others. For example, using IBIR, DeepMutation and CodeBERT contributes with 6.8\%, 0.74\%, and 24.06\%, on the subsuming mutants of PiTest.}}
\label{fig:RQ5-subsumption_relation}
\end{subfigure}
\vspace{-0.8em}
\caption{\small RQ3: Objective comparison between the techniques.}
\label{fig:RQ5-tools-comparison}
\vspace{-0.8em}
\end{figure*}

Figure~\ref{fig:RQ5-tools-simulation} depicts the results of the objective comparison of the 4 techniques. 
Based on these results, we can conclude that the mutants created by PiTest are the strongest ones since a test suite that kills them also kills almost every mutant created by other tools. We can observe similar phenomena in the case of mutants created with CodeBERT. The majority of the mutants seeded by other tools will be killed but remain ~5\% of mutants from IBIR and PiTest that require further analysis. 
We can also observe that test suite designed by mutants from IBIR and DeepMutation will not be very effective in killing PiTest and CodeBERT mutants, leaving us with around ~50\% of the unidentified mutants.

Furthermore, Figure~\ref{fig:RQ5-revailing_bugs} summarises the percentage of faults detected by a test cases selected to kill all the mutants for each tool. 
We depict in gray and green boxes descriptive statistics in terms of median and mean, respectively, for each box plot. 
Thus, we can confirm the earlier findings, where we observed that the test suite designed by PiTest mutants are the strongest ones, being able to detect most real faults (100\% according to median value). 
In the case of test suites designed by mutants from CodeBERT, we can see a trend most similar to PiTest. On average, they can reveal 52.6\% of real faults, which is 1.36\% less than PiTest's average value—at the same time, reporting a median value of 80\% of the real faults. 
On the contrary, we can observe that a test suite designed to identify all mutants from IBIR and DeepMutations shows a significant difference in revealing a fault. In the case of IBIR, we observe that the median value is 0 but the test suitse designed to kill all IBIR mutants can identify on average ~36\% of faults.

\begin{tcolorbox}[
    standard jigsaw,
    opacityback=0,
    left=0pt,right=0pt,top=0pt,bottom=0pt, arc=0pt,
    boxrule=0.4pt,leftrule=1.5pt
]
PiTest and CodeBERT are the strongest techniques capable of almost subsuming IBIR and DeepMutation.
\end{tcolorbox}

Figure~\ref{fig:RQ5-subsumption_relation} depicts the subsuming relation between an observed tool (axis x) and each corresponding tool (pairs).
By observing the relation for each tool separately, we can notice that mutants from PiTest subsume almost every mutant from DeepMutation, and around $\sim$10\% and $\sim$20\% of mutants from IBIR and CodeBERT, respectively, cannot be subsumed by PiTest mutants. This observation follows the same trend for all mutants. Meaning that if we analyze mutants from CodeBERT, we will subsume the majority of mutants from DeepMutation, but will lose nearly 20\% of mutants from IBIR and around 80\% of PiTest generated mutants. 
Mutants from IBIR subsume the majority of the mutants from DeepMutation as well, advocating their subsumption weakness. While on the other side more than 80\% of the subsuming mutants from CodeBERT and PiTest cannot be subsumed.
In the case of mutants from DeepMutation we can see the same trend as before, not subsuming almost any mutant from the other tools. 
To summarise, the robust set of mutants from PiTest can be further complemented with some mutants from CodeBERT, to improve mutation testing capabilities.\\

\begin{tcolorbox}[
    standard jigsaw,
    opacityback=0,
    left=0pt,right=0pt,top=0pt,bottom=0pt, arc=0pt,
    boxrule=0.4pt,leftrule=1.5pt
]
CodeBERT can complement PiTest by offering more than 20\% subsuming mutants to the PiTest subsuming mutant set. 
\end{tcolorbox}

\begin{figure*}[!htb]
\vspace{-0.65em}
\begin{subfigure}[b]{0.33\textwidth}
\centering
\includegraphics[width=\textwidth]{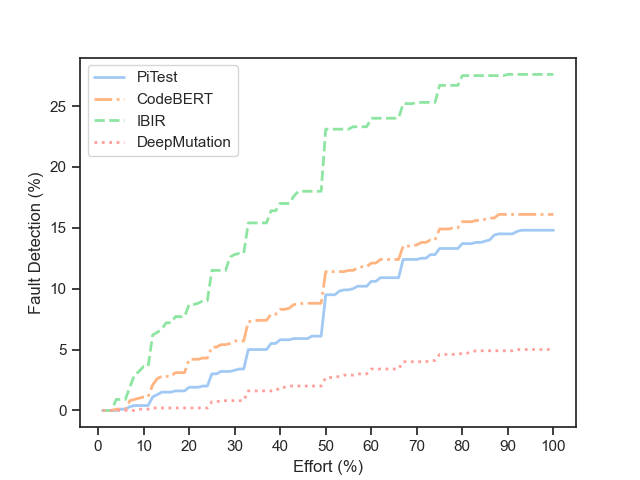}
\vspace{-1.4em}
\caption{{\small DeepMutation.}}
\label{fig:RQ3-Cost-effectiveness-DeepMutation}
\end{subfigure}
\hfill
\begin{subfigure}[b]{0.33\textwidth}
\centering
\includegraphics[width=\textwidth]{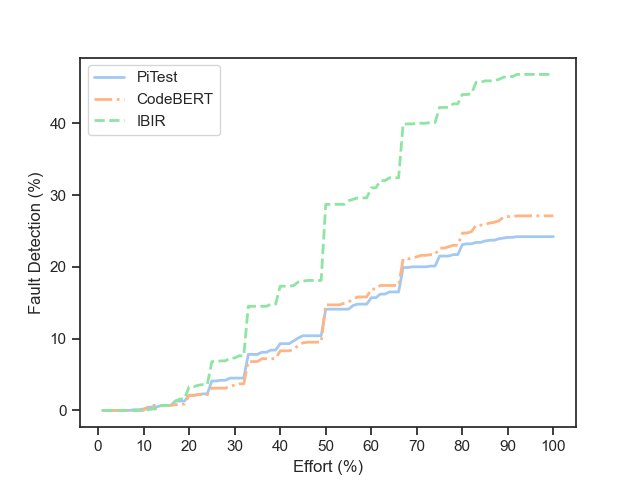}
\vspace{-1.4em}
\caption{{\small IBIR.}}
\label{fig:RQ3-Cost-effectiveness-Ibir}
\end{subfigure}
\hfill
\begin{subfigure}[b]{0.33\textwidth}
\centering
\includegraphics[width=\textwidth]{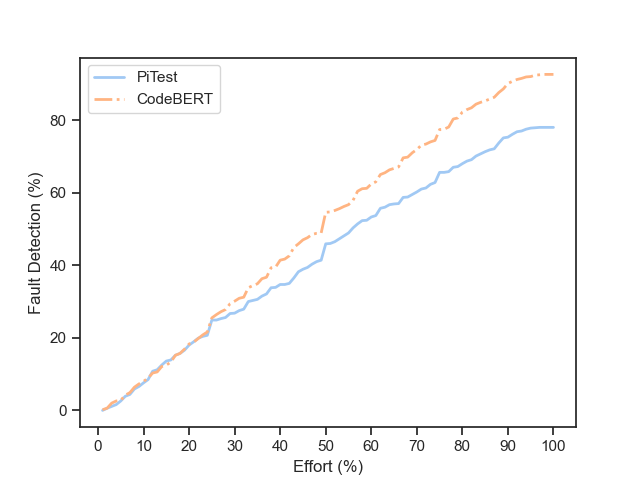}
\vspace{-1.4em}
\caption{{\small CodeBERT.}}
\label{fig:RQ3-Cost-effectiveness-CodeBert}
\end{subfigure}

\begin{subfigure}[b]{0.33\textwidth}
\centering
\resizebox{\columnwidth}{!}{
{\small
\begin{tabular}{l|r|r|r|r|r}
\toprule
\textbf{\% mutants analysed} & 10\% & 20\% & 50\% & 75\% & 100\% \\
\midrule
PiTest - DeepMutation & 0.74 & 0.78 & 0.76 & 0.76 & 0.78 \\
CodeBERT - DeepMutation & 0.79 & 0.80 & 0.85 & 0.85 & 0.85 \\
IBIR - DeepMutation & 0.82 & 0.88 & 0.91 & 0.92 & 0.92 \\
\bottomrule
\end{tabular}
}}
\vspace{-0.1em}
\caption{\small{$\hat{A}_{12}$. Fault detection when number of tests is determined by DeepMutation mutants. IBIR ($\sim$25\%) detects significantly more faults than related tools ($\sim$5\%-$\sim$15\%) under same effort.}}
\label{fig:A12-DeepMutation}
\end{subfigure}
\hfill
\begin{subfigure}[b]{0.33\textwidth}
\centering
\resizebox{\columnwidth}{!}{
{\small
\begin{tabular}{l|r|r|r|r|r}
\toprule
\textbf{\% mutants analysed} & \textbf{10\%} & \textbf{20\%} & \textbf{50\%} & \textbf{75\%} & \textbf{100\%} \\
\midrule
PiTest - IBIR & 0.65 & 0.54 & 0.42 & 0.37 & 0.33\\
CodeBERT - IBIR & 0.50 & 0.48 & 0.39 & 0.36 & 0.32\\
\bottomrule
\end{tabular}
}}
\vspace{-0.1em}
\caption{\small{$\hat{A}_{12}$. Fault detection when number of tests is determined by IBIR mutants. IBIR ($\sim$50\%) detects significantly more faults than related tools ($\sim$25\%-$\sim$30\%) under same effort.}}
\label{fig:A12-IBIR}
\end{subfigure}
\hfill
\begin{subfigure}[b]{0.33\textwidth}
\centering
\resizebox{\columnwidth}{!}{
{\small
\begin{tabular}{l|r|r|r|r|r}
\toprule
\textbf{\% mutants analysed} & \textbf{10\%} & \textbf{20\%} & \textbf{50\%} & \textbf{75\%} & \textbf{100\%} \\
\midrule
PiTest - CodeBERT & 0.44 & 0.49 & 0.44 & 0.43 & 0.42\\
\bottomrule
\end{tabular}
}}
\vspace{-0.1em}
\caption{\small{$\hat{A}_{12}$. Fault detection when number of tests is determined by CodeBERT mutants. CodeBERT ($\sim$90\%) detects more faults than related PiTest ($\sim$80\%) under same effort.}}
\label{fig:A12-Codebert}
\end{subfigure}
\vspace{-2em}
\caption{\small RQ3: Cost-Effectiveness Comparison.}
\vspace{-1.2em}
\label{fig:RQ3-Cost-effectiveness}
\end{figure*}

Finally, in Figure~\ref{fig:RQ3-Cost-effectiveness}, we plot the results of our cost effort-analysis, in particular real fault detection in terms of a number of analyzed tests. 
Precisely, Figure~\ref{fig:RQ3-Cost-effectiveness-DeepMutation} shows that fault detection of DeepMutation is very low ($\sim$5\%) compared to the fault detection reached by the other tools, when same number of tests are taken (determined my the number of tests required to kill DeepMutation mutants).  Clearly, we can observe that under same effort, IBIR is the best cost-effective technique, with $\sim$25\% of fault detection. 
In fact, Figure~\ref{fig:RQ3-Cost-effectiveness-Ibir} evidences that while the number of tests increase (until all IBIR mutants are killed), the fault detection of IBIR ($\sim$50\%) significantly outperforms fault detection capabilities of PiTest ($\sim$25\%) and CodeBERT ($\sim$30\%). 
Figure~\ref{fig:RQ3-Cost-effectiveness-CodeBert} shows that when same number of tests are designed (to cover all CodeBERT mutants), CodeBERT ($\sim$90\%) detects more faults than PiTest ($\sim$80\%). 
Notice that we do not consider PiTest as baseline, since killing all mutants from PiTest leads to large test suites, all mutants of other tools are killed as well with test suites of same size, then no fault detection differences is observed.
Tables~\ref{fig:A12-DeepMutation}, \ref{fig:A12-IBIR} and \ref{fig:A12-Codebert} report Vargha-Delaney A measure~\cite{VarghaDelaney2000} between the fault detection reached by the different tools, indicating that results are statistically significant. \\

\begin{tcolorbox}[
    standard jigsaw,
    opacityback=0,
    left=0pt,right=0pt,top=0pt,bottom=0pt, arc=0pt,
    boxrule=0.4pt,leftrule=1.5pt
]
IBIR is the most cost-effective tool. When controlling for test size, IBIR detects $\sim$50\% of the real faults, while PiTest and CodeBERT detect $\sim$25\% and $\sim$30\%. 
\end{tcolorbox}

%% file: sections/discussion.tex
\section{Discussion} 
\label{sec:discussion}

\subsection{Sensitivity to program locations}
\label{subsec:limitations}

\begin{figure}[t]
\vspace{-0.5em}
\centering
\begin{subfigure}[b]{0.23\textwidth}
    \centering
    \includegraphics[width=\textwidth]{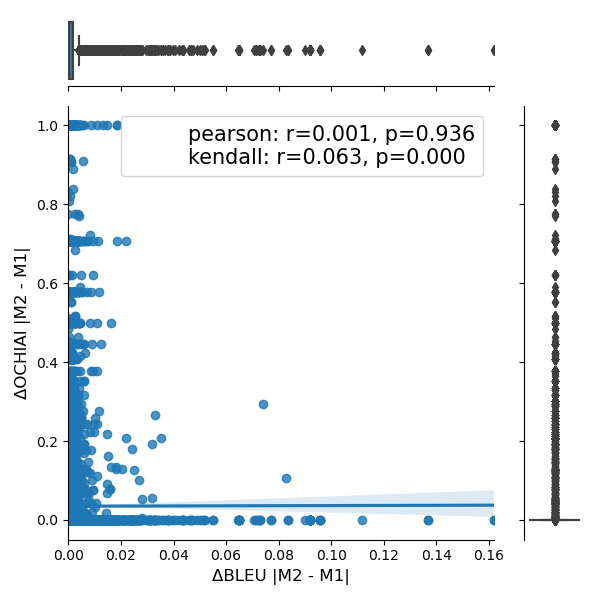}
    \vspace{-1.2em}
    \caption[PiTest - Random lines]%
    {{\small PiTest - Random lines}}    
    \label{fig:PiTest_rq2_mutants_from_random_lines}
\end{subfigure}
\begin{subfigure}[b]{0.23\textwidth}
        \centering
        \includegraphics[width=\textwidth]{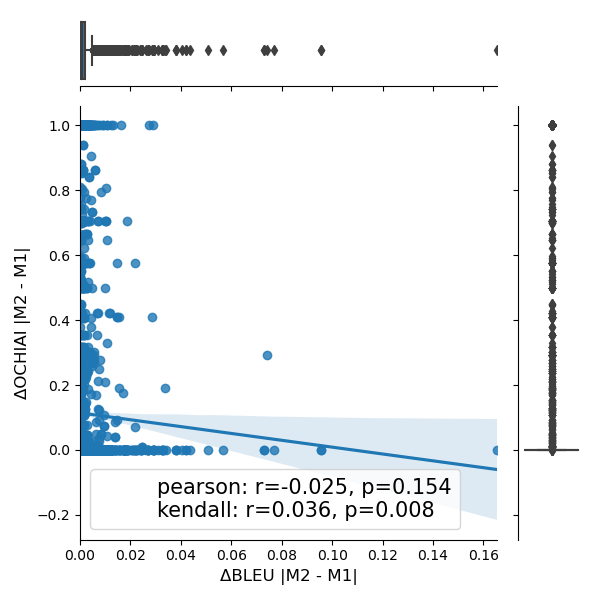}
    \vspace{-1.2em}
        \caption[PiTest - Changed lines]%
        {{\small PiTest - Changed lines}}    
        \label{fig:PiTest_rq2_mutants_from_changed_lines}
\end{subfigure}


\vspace{-1.2em}
\caption{\small  Sensitivity of mutants from the same location. Small syntactic changes lead to diverse semantic changes. }
\label{fig:RQ1-sensitivity}
\vspace{-0.7em}
\end{figure}

One may wonder how sensitive the syntactic and semantic similarity metrics are with respect to the seeded faults' locations. In other words, these metrics may reflect the utility of the locations and not of the faults. Thus, we study the variance of the syntactic and semantic similarity of mutant pairs generated from the same location (we do not consider DeepMutation since it creates only one mutant per method). Figure~\ref{fig:RQ1-sensitivity} records the distances between the syntactic and semantic similarity of mutant pairs, taken from a) the same randomly picked locations and b) from the bug-fixing locations. We observe that while there is almost no syntactic difference between mutants from the same location, the semantic similarity varies significantly. There are a few outliers in which syntactic similarity varies up to 16\% between mutants from the same location. Some PiTest mutations remove a complete line or replace entire boolean conditions by true (as shown in the motivating example), affecting the byte-code generated.  
Figure~\ref{fig:RQ1-sensitivity} includes only plots for PiTest, but a similar trend was observed for all the tools. Please refer to the accompanying website to visualize plots for the other tools.

Overall, these results support the conclusion that there is no link between syntactic and semantic similarity. Interestingly, even small syntactic changes in the same instruction can have a large and diverse impact on the program semantics.


\subsection{Threats to Validity}
\label{subsec:threats-to-validity}

To reduce threats related to external validity, we selected a new and large benchmark of Java language faults that have not been used by previous studies. We though, excluded some faults due to technical reason, making our study with 509 faults from 15 mature open-source real-world projects that are well maintained and tested. Nevertheless, we do not exclude the threat of having different results when conducting the same study on other projects from other domains. As we already discussed, while conducting our experiments, we could not compile or run the tests of all the versions available in Defects4J. 

Other internal threats emerge from the tools' specificity and running configurations, such as the number of mutants to generate, the mutation operators to apply, or source-code locations to mutate. 
For instance, DeepMutation generates one mutant per method,  up to 5 mutants per masked token, IBIR generates mutants only on locations predicted from the bug report by its fault localization step. In contrast, PiTest generates all available mutants in terms of location and mutation operation. To reduce these threats we performed our analysis in a tool basis (without mixing the results from the tools). We also performed a cost-effective analysis by controlling for test suite size to account for the different number of mutants when comparing the tools.   

Unfortunately, we did not manage to compile and run the master-branch~\cite{master-branch-deepmutation} version currently available of DeepMutation. 
We thus, had to recreate the tool from the resources and pre-trained artifacts provided in the repository. 

Additional threat mitigation actions involved the analysis of mutants at different granularity levels (class, method, and location of patch). We also restricted the scope of analysis on the artifacts where the bug-fixes were available, allowing us to consider an original fault from Defects4J as ground truth, isolating it from others. We made sure that all mutants change the same class, or method, or statement, as each of the target faults at a time. Thus, we compare fairly different mutated versions w.r.t. their similarities and distances from the corresponding fixed and faulty version. 


To mitigate any threats from the use of BLEU score when measuring syntactic similarity, we also measured Cosine~\cite{cosine} and Jaccard~\cite{jaccard} similarity coefficients. The results did not showed any significant differences wrt the ones of BLEU scores. Nevertheless, please refer to the accompanying website for additional details on using Cosine and  Jaccard similarity.

Other threats may arise from the test suites used. Since the employed projects are mature and well-tested they should approximate well vital project semantics. 

